\documentclass[acmsmall]{acmart}

\AtBeginDocument{%
  \providecommand\BibTeX{{%
    \normalfont B\kern-0.5em{\scshape i\kern-0.25em b}\kern-0.8em\TeX}}}

\acmDOI{XXXXXXX.XXXXXXX}

\setcopyright{none}

\usepackage{gensymb}
\usepackage{tabularx}
\usepackage{caption}
\usepackage{subcaption}
\usepackage{amsmath}
\usepackage{comment}
\usepackage{wrapfig}
\usepackage[utf8]{inputenc}
\newcommand{\bdj}[1]{{\small\color{red}[Brendan: #1]}}
\newcommand{\candace}[1]{{\small\color{orange}[Candace: #1]}}

\begin{document}

\setlength{\belowcaptionskip}{4pt}
\setlength{\abovecaptionskip}{4pt}
\setlength{\textfloatsep}{8pt}

\title{GazeIntent: Adapting dwell-time selection in VR interaction with real-time intent modeling}

\author{Anish S. Narkar}
\email{anishnarkar@vt.edu}
\affiliation{%
  \institution{Virginia Tech}
  \city{Blacksburg}
  \state{Virginia}
  \country{USA}
  \postcode{24060}
}

\author{Jan J. Michalak}
\email{janmichalak@vt.edu}
\affiliation{%
  \institution{Virginia Tech}
  \city{Blacksburg}
  \state{Virginia}
  \country{USA}
  \postcode{24060}
}

\author{Candace E. Peacock}
\email{peacock.candace@gmail.com}
\affiliation{%
  \institution{Independent Researcher}
  \city{Denver}
  \state{Colorado}
  \country{USA}
  \postcode{80005}
}

\author{Brendan David-John}
\email{bmdj@vt.edu}
\affiliation{%
  \institution{Virginia Tech}
  \city{Blacksburg}
  \state{Virginia}
  \country{USA}
  \postcode{24060}
}

\renewcommand{\shortauthors}{Anish S. Narkar, Jan J. Michalak, Candace E. Peacock, \& Brendan David-John}

\begin{abstract}
The use of ML models to predict a user's cognitive state from behavioral data has been studied for various applications which includes predicting the intent to perform selections in VR. We developed a novel technique that uses gaze-based intent models to adapt dwell-time thresholds to aid gaze-only selection. A dataset of users performing selection in arithmetic tasks was used to develop intent prediction models\,(F1 = 0.94). We developed \textit{GazeIntent} to adapt selection dwell times based on intent model outputs and conducted an end-user study with returning and new users performing additional tasks with varied selection frequencies. Personalized models for returning users effectively accounted for prior experience and were preferred by 63\% of users. Our work provides the field with methods to adapt dwell-based selection to users, account for experience over time, and consider tasks that vary by selection frequency.
\end{abstract}


\received{November 2023}
\received[revised]{January 2024}
\received[accepted]{March 2024}

\maketitle

\section{Introduction}
\label{Sec:introduction}
Virtual Reality\,(VR) and Augmented Reality\,(AR) headsets employ hand controllers and gestures to interact with virtual environments\,(VEs). These interaction methods perform a wide range of functions, such as selection, manipulation, and locomotion. Selection is one of the most common actions performed by users in a VE. There has been a pivot towards controller-free selection in VE using tracked hand gestures~\cite{tian2019analyzing,hirzle2023xr, gazegesture2, yang2019gesture, khundam2015first, spittle2022review}. However, gesture-sensing tools often have a limited range that constrains the space where gestures can be detected, leading to fatigue.  Cameras and other sensors used for hand tracking and gesture recognition are affected by environmental factors such as brightness and occlusion which can lead to inaccurate or missed gestures. These limitations have led to a broad range of hands-free interaction techniques for 3D VEs. In this work, we investigate gaze-based selection that only considers the user's eye movements.

One of the earliest proposals of gaze-based computer interaction was presented in 1990 and considered gaze-based selection for task-relevant objects on a computer monitor~\cite{onlygaze4}. Future research then demonstrated the intuition that selection using gaze is faster than hands~\cite{tanriverdi2000interacting}. However, a major limitation of gaze-only interaction was the likelihood of unintentionally selecting objects the user only meant to glance at. Typically, thresholds of how long a user's gaze must dwell on the object to select have been used to balance selection time and false selections. More recent work have used gaze dynamics to predict user intent to trigger gaze selection based on the model instead of dwell time~\cite{whatdowedo, david2021towards, GazeIntent, LookandPoint}. Prior results from intent modeling are encouraging, however, there is a lack a comprehensive cross-task evaluation when intent models are trained on one task and deployed on another. 

Our primary contribution in this paper is a novel gaze-based selection architecture, \textit{GazeIntent}, that uses a temporal prediction model of interaction intent to scale and adapt dwell-time thresholds for gaze-only selection. Our approach applies a threshold to the model prediction outputs and scales the dwell time for selection to both counteract the influence of false positives from spurious model predictions and enables faster selection than a pre-determined static dwell time. We evaluated our approach in VR by first collecting interaction data from a task performed with a hand controller for initial model training and then demonstrated successful deployment in a range of tasks using only gaze-based selection. We contribute a novel intent model that outperforms prior work with an F1 score of 0.94 and found that our \textit{GazeIntent} interface was (1) preferred over state-of-the-art intent-based dwell selection, (2) accommodated new selection tasks, and (3) was capable of creating personalized models that were preferred over a general model by returning users. 


\begin{figure}[t]
    \centering
    \includegraphics[width=0.96\textwidth]{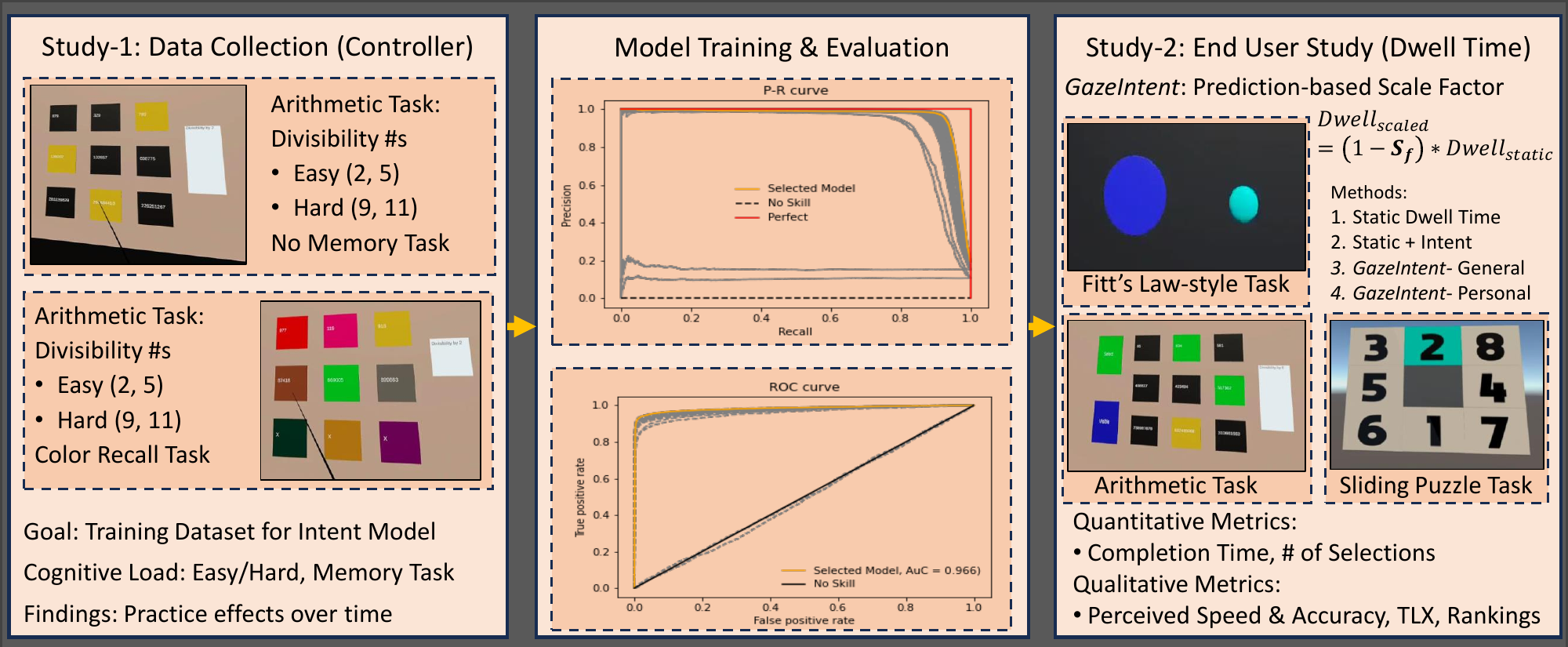}
    \caption{ We conducted a user study to build a comprehensive VR interaction dataset, trained intent prediction models on the dataset, and deployed them with \textit{GazeIntent}, our adaptive dwell time scaling gaze selection method. 
    }
    \label{fig:teaser}
\end{figure}

\section{Related Works and Motivation}

In this section, we discuss relevant work on  dwell time-based gaze selection, applications of gaze-based prediction, and techniques that implement gaze-based intent prediction models for selection.
\paragraph{Dwell Time Gaze-based Selection} 
 A typical gaze-only interaction sets a minimum dwell time for how long a user must look at an object to trigger selection. A survey by Plopski et al. provides deeper discussion of gaze-based interaction systems for extended reality devices~\cite{plopski2022eye}. 
Researchers have explored optimizing this dwell-time threshold to improve the performance of gaze-based selection. For example, OptiDwell~\cite{onlygaze5} uses a reinforcement learning framework to account for changes in user behavior over time and optimized thresholds per user and interface type. The authors performed an end-user study where the subjects interacted with the system and recorded every false gaze selection as it occurred using a keyboard. The authors used the frequency of key logs to show that their proposed model improved selection accuracy by optimizing thresholds over five sessions. 


\paragraph{Gaze-based Prediction} Gaze data has also been used extensively to study the onset of interaction events or changes across tasks. Within perception and attention studies, the temporal dependencies between eye and hand movements are highly correlated across tasks~\cite{hayhoe2003visual}. These findings led researchers to leverage gaze to build inference models of user behavior. For example, gaze-based models have been used to predict when a user is lost during a navigation task in VR~\cite{LostInStyle}. 
The authors observed a drop in predictive performance of an LSTM when the model was deployed in a different VE than the one used to train the model, suggesting that models need to adapt to different tasks and environments. 
 Eye movements along with positional tracking have also been used to predict motion trajectories while walking in VR~\cite{bremer2021predicting}. 
 

Gaze-based models also have predictive properties related to deeper cognitive state of the user. For example, VR studies have shown that models trained on gaze dynamics can predict working memory encoding~\cite{peacock2022gaze_wm}. Participants performed a guided search task in a virtual apartment using visual cues\,(arrows) and were asked to recall the objects they were guided to. A logistic regression model was able to successfully predict which gaze behaviors occurred while encoding the target objects. A follow-up study evaluated the model trained on the first experiment in a new environment with new users. 
Their results revealed a subset of features that correlated with working memory encoding, i.e., object recall, even when trained on different users and environments. Similar models and feature sets have also been used to predict when a user experiences a system error in VR selection~\cite{peacock2022gaze_errors}. The models were deployed to predict when a user noticed an erroneous selection by synthetically injected errors in the system. Taken together, these findings suggest that gaze dynamics are sensitive to both cognitive states and interaction events; and influenced our study design to include interaction modeling under varying cognitive states and tasks.  


\paragraph{Gaze-Intent based selection} Bednarik et al. proposed to use ML models for intent prediction in a 2D sliding puzzle game~\cite{intentmodel3}. The authors predicted intent using several features, such as duration, gaze velocity, and path length during saccade and fixation events. 
The model was evaluated offline and was not deployed in real time. 
Features derived from gaze dynamics have been used to predict the intent of users to select an object in a virtual pantry\,(in VR) using an ML model~\cite{david2021towards}. The authors showed that gaze dynamics, which are less susceptible to spatial accuracy errors in the eye tracker, can be used to predict intent. Features like fixation detection, gaze velocity, and gaze velocity specifically during fixation were some of the most impactful features. The main limitation of this work was that the model was not deployed in real-time and performance metrics for precision were low\,(AUC-PR=0.12). The authors had shown that the top-ranked features were not the same between the users, suggesting that fine-tuning or personalizing a model for each individual could create more accurate intent models.

The most recent approach for model-based gaze selection paired intent models with dwell-time selection in a two-step process~\cite{GazeIntent}. 
First, the current dwell time should exceed the static threshold for object selection. Then, the output of an intent prediction model must also produce a positive classification to confirm the selection. 
The Isomoto intent model used ~127 features in a 2-second input window and random forests for classification. A major limitation of this work was that the users needed to look at an object for the entire duration of the dwell-time threshold. While this approach limits false positives, it creates a lower bound on the speed of interaction. The dwell time must hit this lower bound and then wait until the intent model also indicates selection, affecting user experience during tasks with high selection frequency. 
Based on these works, we can summarize the \textbf{key limitations} of current approaches as follows: 

\textit{Different Selection Frequency}: A model trained on a task with slow interaction frequency might not generalize well on a task with a faster selection frequency. The selection frequency affects the balance between positive and null class labels. Methods to account for class imbalances during training can mitigate this effect at training time. However, it is necessary to deploy and evaluate the generalization of models with tasks of differing selection frequencies. Most of the existing works did not deploy their evaluations in real time~\cite{peacock2022gaze_wm, GazeIntent, david2021towards}. In our evaluation, we trained our model on a division task ranging from one to nine selections per minute and deployed on tasks where the interaction frequency ranged from 20 to 40 selections per minute. We then tested whether users could subjectively perceive a difference in their interactions between models.

\textit{Task Experience:} The selection speed of users increases with practice, and only the optimized dwell-time threshold approach accounts for these changes~\cite{onlygaze5}. 
The typical process of splitting training and testing data does not maintain the temporal ordering of user behavior as they become familiar with the task. We implemented a forward chaining training-validation approach to progressively train the model over time. 

\textit{Different User States}: Deployed ML models are typically validated in a controlled environment. When models are deployed in the real world, it is difficult to anticipate a user's cognitive state and whether the trained model will generalize to their current behavior. 
We varied cognitive load during collection of our training dataset and evaluated models with subsets of the collected data.

\textit{Adaptibility to New Users:} The ability of intent prediction models to adapt to user behavior not present in the training dataset is crucial to real-world deployment. We evaluated our interface with new and returning users.

\section{Training Dataset \& Model Development} In this section, we describe the protocol to collect our VR interaction training dataset\,(Sec.\,\ref{sec:study1-collection}), define the pipeline for processing and extracting features from gaze data\,(Sec.\,\ref{sec:-study1-datapipe}), illustrate our model architecture\,(Sec.\,\ref{sec:study1-training}), and present offline model performance results\,(Sec.\,\ref{sec:study1-results}).

\subsection{Data Collection}
\label{sec:study1-collection}
We conducted an initial user study to train intent models offline and find an optimal model. The study was approved by our Institutional Review Board\,(IRB) and subjects were paid \$20 for a 90-minute session. We collected data from eighteen subjects (7 Females, 8 Males, 1 Non-Binary; Mean Age = 25$\pm$ Std. Dev. of 6 Years). 
Subjects were shown three numbers at a time and had to determine which one was divisible by a given divisor (e.g., which of 70, 81, and 91 are divisible by 9). We refer to this divisor as the divisibility number\,(DN). We considered four different DNs\,(2, 5, 9, 11) and explained the mathematical rules for testing divisibility with these numbers before data collection.\footnote{For example, a number is divisible by five if it ends with a five or a zero, while a number is divisible by nine if the sum of all digits ends with a nine.} The DN task varied in difficulty, as testing for divisibility by 2 and 5 is easier than 9 and 11. 

The VR environment was developed using Unity 2021.3.23f1 and deployed on the Oculus Quest Pro VR headset. Eye-tracking data was logged at a sampling rate of approximately 66Hz\,(avg. time deviation between logged samples was 15ms). We logged gaze direction using a Gaze-In-World\,(GIW) coordinate frame~\cite{diaz2013real}, in line with prior VR intent modeling works~\cite{david2021towards}. 
Numbers were displayed on a 3x3 grid of 3D cubes, with only one row visible at a time\,(Fig.\,\ref{fig:teaser}, Left). Each cube displayed a randomly generated number. The first, second, and third rows had numbers with three, six, and nine digits, respectively. In each row, there was only one correct number divisible by the DN. The numbers were determined randomly for each grid. Subjects were asked to make selections using a hand controller by pointing the controller ray at the cube and pulling the trigger. All participants were right-handed. 

At the start of the study, numbers were not visible to the subjects. Subjects were asked to point the controller toward the floor and pull the trigger to make a row visible. Subjects then had to identify and select the cube containing their answer, changing the color to yellow. To reset the subject's hand position after each selection, they were asked to point the controller at the floor again and pull the trigger to make the next row appear. Each trial ended after the third row was completed. We defined one round as three consecutive trials with a distinct DN and introduced an additional memorization task for the second and third trials. After completing the first trial of each round, the cubes selected by the subject were colored randomly from a set of 12 colors. Subjects were provided the 12 color values and their names prior to data collection. Subjects were asked to memorize the sequence of colors from each row\,(top to bottom) for recall at a later time. Once the subject was ready to proceed, they selected a prompt to begin the next trial. 

Subjects then performed two more full selection trials. To increase the difficulty in color memorization, random colors were assigned to all of the cubes in these trials. After finishing the round, subjects were asked to verbally recall the three colors they had memorized which were logged by the experimenter. Each subject performed eight rounds\,(DNs = 2, 5, 9, 11, 2, 5, 9, 11) per block for three blocks. A mandatory five-minute break was enforced between each block.

\subsection{Data Pipeline}

Our data pipeline consists of four steps: signal processing, event detection, feature extraction, and data splitting. 

\paragraph{Signal Processing}\label{Sec:Pre-Processing} The first step in the data pipeline is to compute the instantaneous velocity between successive gaze samples. The displacement was computed as the arccosine of the dot product between 3D vectors and velocity was derived by division using $\Delta t$. We then filtered the raw velocity to measure gaze dynamics. 
We applied the Savitsky-Golay\,(SG) filter to all velocity values~\cite{savitzky1964smoothing, Holmqvist}. The SG filter performed convolutions over adjacent data points using a low-order polynomial. Velocities over 800\degree /s were considered outliers~\cite{dowiasch2015effects} and were removed from the signal and replaced via linear interpolation. We also computed horizontal and vertical components of velocity and performed another derivative to measure acceleration. 


\paragraph{Event Detection}
Events were detected from gaze velocity using the I-VT algorithm~\cite{goldberg}. A fixation was detected when the velocity was below 30\degree/s with a minimum duration of 30 msec; while a saccade was detected when the gaze velocity exceeded 70\degree/s with a minimum and maximum duration of 0.02 sec and 0.2 sec respectively~\cite{nystrom2010adaptive, olsen2012identifying, olsen2012tobii}. 

\paragraph{Feature Extraction}
Our proposed gaze model combined continuous features directly from gaze signals and discrete features corresponding to each event. Continuous features were generated during the signal processing stage of the data pipeline. 
These features included displacement, velocity, and acceleration of gaze signals in horizontal, and vertical directions. It also included mean, median, mode, standard deviation, skew, and kurtosis of velocity in these directions.
 Discrete event-based features are defined based on fixations and saccades. The event-derived features include an event boolean for each sample (to indicate the presence or absence of fixation or saccade), duration of each event, dispersion, and path length during fixations. A detailed list of all the features used is provided in the Supplementary Material. The low-level event-based features were generated as described in~\cite{david2021towards} and are adapted from models of biometric recognition~\cite{george2016score}. There were 45 fixation, 44 saccade, and 9 continuous features in total. Discrete features were appended to the continuous feature values at the sample after the event ended. For all other samples, the discrete features are set to zero.

\paragraph{Data Windowing}
The input to the model was a sliding window of multivariate time series data. Windows were defined by size\,(number of samples it spans) and the window overlap\,(number of samples included from the prior window). The subjects performed other actions beyond the division tests, such as reading instructions and navigating between different scenes in the study. Only gaze data from when users were actively performing the divisibility tests was used for training the ML model. The optimal value for both parameters was identified during training~\ref{sec:study1-training}. 

\paragraph{Data Labeling}
 To label the ground truth intent data we needed to identify the temporal moment when subjects decided to select a cube. We followed the label-assigning method inspired by Lost In Style~\cite{LostInStyle}, in which all feature windows that end within a set amount of time from the user pulling a trigger were labeled as the positive class. The rest of the windows were labeled as negative. 

 \label{sec:-study1-datapipe}
\begin{figure}[t]
    \centering
    \includegraphics[width=\textwidth]{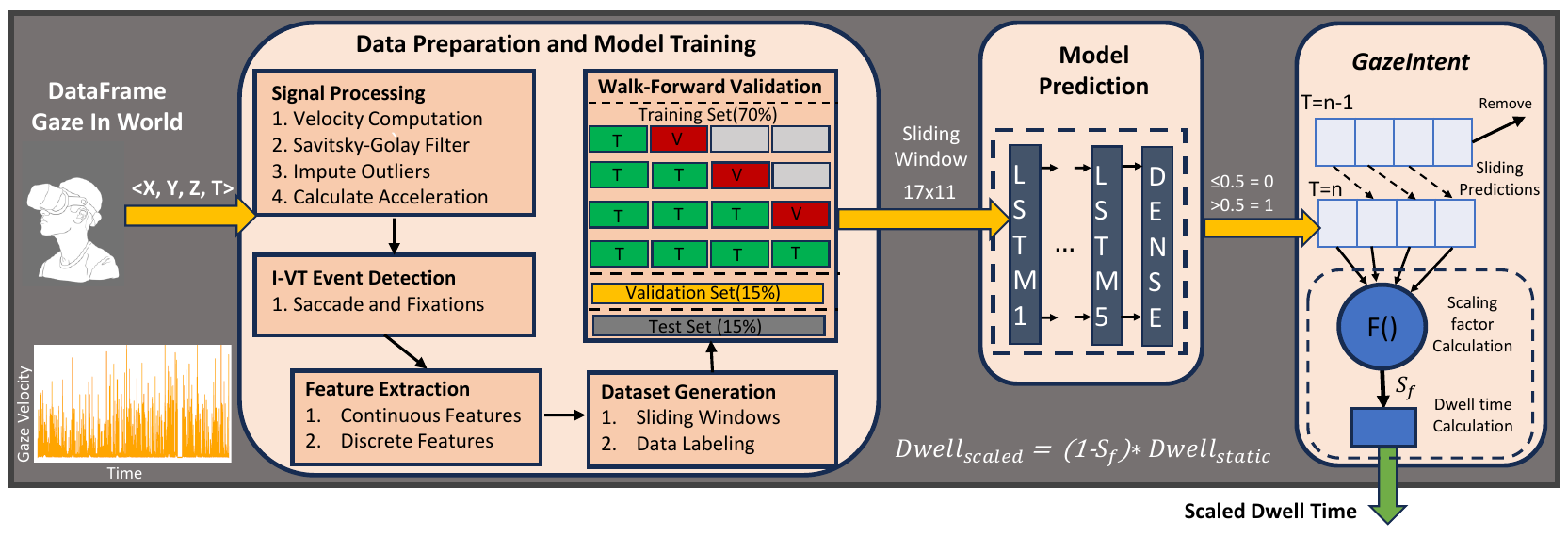}
    \caption{Intent modeling training process and \textit{GazeIntent} system architecture for scaling dwell time. Raw GIW data is processed to create a dataset and train a temporal intent model. The intent model is deployed by maintaining sliding windows of the last four predictions to compute a Scaling Factor $S_f$\,(Eq.\ref{eq:scalingfactor}). The $S_f$ is then used to dynamically adjust the dwell-time threshold for selection.}
    \label{fig:systemarchitecture}
\end{figure}
 
\subsection{Model Training and Selection}
\label{sec:study1-training}
We trained a 5-layer long short-term memory\,(LSTM) model that consists of 512, 256, 128, and 64 units of LSTM using TensorFlow~\cite{hochreiter1996lstm}. The output of the final LSTM layer is passed through a dense layer with sigmoid activation to output a scalar value between 0 and 1. The loss function was binary cross entropy with an ADAM optimizer~\cite{kingma2014adam}. The datasets were imbalanced, with the ratio of negative to positive class being being 9:1. Class weights based on the proportion of samples were assigned to account for imbalance~\cite{guo2008class}. 

We trained and evaluated models to optimize the system configurations and model structure\,(hyperparameters). We trained each model for 200 epochs. We employed early stopping to stop the training (after 20 epochs) when improvement plateaued~\cite{yao2007early}. System configurations include the optimal number of samples in the sliding window, the size and order of the SG filter, the ground truth label duration, and the overlap count.

\textbf{Stage-1:} In the first round of model development, we trained 288 models.  Each model configuration took $\sim$20 minutes to train on an Nvidia RTX 3080 Ti and was based on four sliding window lengths $\langle$13,15,17,19$\rangle$, six combinations of SG filter order and size $\langle$ (1,2), (7, 9,11)$\rangle$, four positive label intervals $\langle$0.5,0.75, 1,1.25$\rangle$, and three values of overlap count $\langle$0,1,2$\rangle$.  The goal of this stage was to find the best combinations of model configurations when considering the full set of 98 features. Data were split temporally, with the first 70\% of the data session for each user included in the training set, the next 15\% in the validation, and the final 15\% being in the test set. To avoid data leakage between users and temporal splits, data normalization was applied individually within each user's data splits. 
At the end of this stage, we identified the top 50 models based on the F1 score. 

\textbf{Stage-2:} In this stage, we further refined our models to be less prone to overfitting. Typically, K-fold cross-validation or LOOCV are used to reduce the chances of overfitting. However, random sampling in time series data might result in a scenario where we train a model to predict past events on future samples, which does not capture changes in selection behavior over time. Thus, we employed the walk-forward validation technique, which has been applied in other temporal domains such as economic and financial forecasting~\cite{kaastra1996designing, siami2018comparison}. We split the training data from the first stage into five equal blocks temporally. An illustration of this technique is shown in Fig.~\ref{fig:systemarchitecture}\,(Model Training). In the first cycle of training, we trained the model on the first two blocks and validated it on the third block. In the second cycle, we trained the model from the first cycle on the first three blocks and validated on the fourth, while in the third cycle, we trained on the first four blocks and validated on the fifth block. 
We started with the 50 best models from Stage 1 and trained the model from scratch using the walk-forward validation method. After the final cycle, we identified the top ten models based on the F1 score. 

\textbf{Stage-3:} The objective of this stage was to find the optimal set of features for the model. In this stage, we first determined the features that most impacted the model prediction using Shapley analysis and formed feature subsets using only optimal features. The top models were then re-trained to identify the optimal model and feature set. 

\subsection{Results}
\label{sec:study1-results}
The goal of our evaluation was to identify the most impactful gaze features for predicting intent, identify optimal intent model parameters for future deployment, and determine whether cognitive load had an impact on model performance. Figure\,\ref{fig:teaser}\,(Middle) presents precision-recall\,(PR) and receiver operating characteristic\,(ROC) performance curves for the models trained in this section. The majority of models performed well\,(F1 $\geq$ 0.9) compared to the best performing prior gaze-based selection intent model\,(F1 = 0.8)~\cite{GazeIntent}.

\paragraph{Shapley Analysis} Shapley analysis measures the impact each feature has on model predictions~\cite{SHAP1}. To identify important features, we applied this analysis to the top ten models from Stage-2. Table~\ref{tab:shapley} shows that the continuous features had stronger influence on predictions than event-based features. The mean influence of continuous features on the prediction was 1.644 $\pm$ 0.89, while the mean influence of event-based features was 0.05 $\pm$ 0.009. 

To further explore the performance of feature sets we created the following feature sets for an ablation study: 
 \begin{itemize}
     \item Set-1: Only continuous features
     \item Set-2: Continuous + Boolean indicating the presence and absence of the event
     \item Set-3: The Top-30 most impactful features from SHAP.
     \item Set-4: Only Event-based features
 \end{itemize}
 
To corroborate the results of the Shapley evaluation, we retrained the top-ten models on all four new feature sets. 

\begin{table}[]
\centering
\caption{Influence of features selected for model deployment. Continuous features are primarily affecting the model prediction  }
\label{tab:shapley}
\begin{tabular}{lll|lll}
\hline
Rank & Feature                 & Influence & Rank & Feature             & Influence \\ \hline
1        & Absolute Acceleration   & 2.7       & 7        & Horizontal Velocity & 0.6       \\
2        & Displacement            & 2.4       & 8        & Absolute Velocity   & 0.5       \\
3        & Vertical Displacement   & 2.2       & 9        & Vertical Velocity   & 0.4       \\
4        & Horizontal Displacement & 2.2       & 10       & Fixation Boolean    & 0.12      \\
5        & Vertical Acceleration   & 2.0       & 11       & Saccade Boolean     & 0.1       \\
6        & Horizontal Acceleration & 1.8       &          &                     &           \\ \hline
\end{tabular} 
\end{table}

\paragraph{Effect of Cognitive Load:} We induced cognitive load using two factors, color memorization and arithmetic complexity. We generated 4 test subsets from our test set based on the presence and absence of the cognitive load from each of these two factors. We tested the performance of our top 50 models on each of these four subsets. Table\,\ref{tab:model-perf}\,(Left) compares the results of top-50 models on each of the subsets and on a general test set that consisted of all these subsets.

\begin{table}[]
\caption{Left: The effect of induced cognitive load on model performance across test sets. (Right) The effect of features on model performance for the full feature set.} 
\label{tab:model-perf}
\begin{tabularx}{0.42\textwidth}{ll}

\hline
Test Dataset          & F1-Score \\ \hline
No math Load                  & $0.963 \pm 0.003$   \\
Math Load       & $0.919 \pm 0.005$     \\
No Memory      & $0.927 \pm 0.005$      \\
Memory & $0.926 \pm 0.006$      \\
All     & $0.927 \pm 0.001$       \\ \hline
\end{tabularx} 
\begin{tabularx}{0.5\textwidth}{ll}

\hline
Feature Set                   & F1-Score \\ \hline
All features                  & $0.927 \pm 0.003$   \\
Top-30 Shapley Features       & $0.925 \pm 0.009$     \\
Only Continuous Features      & $0.924 \pm 0.012$      \\
Continuous + Boolean Features & $0.927 \pm 0.008$      \\
Only Event-based features     & $0.268 \pm 0.050$       \\ \hline
\end{tabularx}
\end{table}

\paragraph{Optimal Feature Set:} The models' performance indicated no practical difference in the performance of models trained on complete feature sets and models trained using only continuous, continuous + event booleans, and the top 30 most impactful features. The models trained using continuous and event booleans marginally outperformed the other feature sets. The models trained only on event-based features performed worst overall. 



\paragraph{Model performance under load:} We performed two paired samples t-tests to compare the performance of models trained with and without induced cognitive load. We compared F1 model performance on subsets within each dimension of cognitive load. The t-tests found no significant difference in model performance in both cases\,(\textit{p} > 0.05). 


\textbf{Based on the F1 score, we identified the optimal model parameters of: window size of 17 samples, window overlap of 2, SG filter of length 11 and order-1, and ground truth label interval of 1 second.} 

\section{Real-Time Model Deployment}
The intent prediction model indicated if the features of an input sliding window corresponded to an interaction event or not. Our proposed \textit{GazeIntent} architecture leverages this prediction to scale the necessary dwell time for selection. In this section, we present our approach to mapping intent predictions to an adaptive dwell time scaling factor.  

\subsection{Dwell Time Scaling}
A linear scaling mapping function is used to compute the dwell time necessary for selection as

\begin{equation}
  \label{eq:dwellpred}
    Dwell_{scaled} = (1-S_f) * Dwell_{static},
\end{equation}
where $S_f$ is the Scaling Factor and is determined by the prior intent model predictions. The $S_f$ determines the ratio by which the dwell time is reduced. The range of this factor is $0<=S_f<=1$. When $S_f = 0$ the dwell time is not scaled and is equal to the static dwell time threshold. When $S_f=1$ the dwell time becomes 0, leading to an immediate selection. 

The $S_f$ is defined as a linear combination of prior model prediction values. 
In line with our ground truth labeling duration of one second and input window size of 17 samples, we calculated the number of input windows generated in one second. With a frequency of 66 Hz and an overlap count of 2 samples, four input windows span one second. Thus, we consider the last four predictions to calculate the scaling factor as 
\begin{equation}
  \label{eq:scalingfactor}
    S_f = P_n*T_{n} + P_n * P_{n-1}T_{n-1} + P_n * P_{n-1}*P_{n-2}*T_{n-2} + P_n * P_{n-1}*P_{n-2}*P_{n-3}*T_{n-3},
\end{equation}

where $P_n$ is a binary prediction at $time=n$; and $T_{n}, T_{n-1}, T_{n-2}, T_{n-3}$ are constants where $T_k$ indicates the weight of prediction at $time=k$. The constant threshold values can be fixed based on task or user preferences and balance the contributions of previous predictions to the current scaled dwell time. 
We enforced constraints that each threshold value $T$ was between zero and one, and that the sum of all values must be equal to one to ensure that scaled dwell time is not negative. Threshold values affected how quickly an object was selected. If the threshold for the latest prediction\,($T_{n}$) was high, then scaled dwell time would become faster quickly; while a large threshold for the earliest prediction\,($T_{n-3}$) would scale the threshold gradually.

 \subsection{Implementation Details:} Our intent model was deployed within Unity in real time using the Barracuda inference library. 
We maintained a raw data array to aggregate gaze samples, an intermediate array of feature frames to input to the intent model, and the last four model prediction values for computing the current $S_f$. The dimension of the intermediate array was 17x11 to match the model's input window size. 
Once enough new samples were stored to define a new input window, the displacement, velocity, and acceleration were computed and processed via filtering and outlier handling as mentioned in the previous section. We classified each sample as fixation, saccade, or neither to generate the two event boolean signals and fused them with displacement, velocity, and acceleration continuous features. The fused feature data was then added to the intermediate array which was input to the prediction model. The most recent model prediction was added to the list of predictions and the oldest prediction was removed. 

\subsection{End User Study}
We conducted a second user study to evaluate the performance of our model during real-time deployment. The study was approved by our IRB and subjects were paid \$20 USD for a 90-minute study. We recruited 16 subjects (5 Females, 10 Males, 1 Non-Binary, Mean Age: 25 $\pm$ Std. Dev. 5 Years), with eight of these subjects returning from the first user study. We did not include color memorization in our deployment tasks since we did not find conclusive evidence for the impact of memory on model performance~\ref{sec:study1-results}. The tasks and the methods were all deployed on Oculus Quest Pro and Unity 2021.3.23f1.      

\subsubsection{Tasks}
\label{sec:tasks}
We evaluated the following four tasks: 
\begin{itemize}
    \item \textbf{Circle Selection/Fitt's Law-style task:} Participants were asked to select a circle of random size and distance with their eyes. They were asked to select as many circles as possible in 60 seconds.
    \item \textbf{Arithmetic task, same stimuli:} Participants performed the same grid divisibility tests with DNs from Study 1.
    \item \textbf{Arithmetic task, different stimuli:} Participants also performed the grid divisibility tests for different numbers (3,4,6,12). This task was included to verify that prior modeling results were generalizable to task variations within the same environment. 
    \item \textbf{Sliding Puzzle Task:} Participants solved a sliding 9-tile puzzle. In this puzzle, participants needed to rearrange numbers from 1-8 in increasing order from top to bottom by moving them around using an empty block. This task was included as a prior application of gaze-based intent modeling~\cite{whatdowedo}. 
\end{itemize}

\subsubsection{Methods}
We deployed four gaze-only object selection methods: 
\begin{itemize}
    \item \textbf{Static Dwell Time\,(Baseline):} To select an object, the dwell time needs to surpass a static value.
    \item \textbf{Static + Intent\,(S + I):} We did not have access to some of the variables used by Isomoto et al\,\shortcite{GazeIntent}, so instead of deploying their exact model we deployed a model based on the framework laid out by them. We waited for the dwell on a particular object to surpass a static value and then used our intent model to predict intent. If the prediction was positive, then the corresponding object was selected. We imposed an upper dwell time limit of double the static threshold for this method in the event of many negative predictions by the model. 
    \item \textbf{\textit{GazeIntent} General Model\,(GI-G):} This is our proposed \textit{GazeIntent} dwell scaling system. The model driving the intent prediction is trained on the data from all users.
    \item \textbf{\textit{GazeIntent} Personalized Model\,(GI-P):} This is our proposed \textit{GazeIntent} dwell scaling system, but it is first trained without the target user's data. The model is then fine tuned on the target user's data for 200 epochs. We could only deploy this method for subjects returning from the first user study.
\end{itemize}

The presentation order of selection methods were counterbalanced across subjects. We were not exploring behavior between selection tasks and therefore did not counterbalance task order. Participants first performed the Fitt's Law-style task, followed by the Arithmetic tasks, 
and then the Sliding Puzzle task. 

\subsubsection{Pilot Study}
We conducted a pilot study to identify an appropriate static dwell time for each task in the end-user study. Several participants and the authors completed each task across a range of dwell times and indicated their preferences. The dwell times identified for the Circle, Arithmetic, and Sliding Puzzle tasks were 0.3 sec, 1.5 sec, and 1.2 sec respectively. Next, the same users experimented with three configurations of threshold values: (1) $T_{n} < T_{n-1} < T_{n-2} < T_{n-3}$, (2) $T_{n} = T_{n-3} < T_{n-2} = T_{n-3}$, and (3) $T_{n} < T_{n-3} < T_{n-1} << T_{n-2}$. The third configuration was preferred by all participants. We deployed threshold values of $T_{n}=0.05$, $T_{n-1}=0.1$, $T_{n-2}=0.15$, $T_{n-3}=0.7$.

\subsubsection{Metrics}
\label{sec:metrics}

\paragraph{Quantitative}
\label{sec:quant_metrics}
We computed quantitative metrics to capture differences in selection performance across the four methods. The metric for the Fitt's Law-style and Sliding puzzle tasks was the total number of selections made\,(\# of Selections). Tracking the total number of selections is linked to Fitt's Law-style tasks goal of selecting as many circles as possible within the fixed time limit. Likewise, the Sliding Puzzle Task required participants to act quickly to solve the puzzle within three minutes. Most participants did not finish the puzzle in time. The Arithmetic task was evaluated using time-to-completion\,(TTC) to evaluate how quickly participants could complete the divisibility test for each grid.  

\paragraph{Qualitative}
\label{sec:qual_metrics}
Subjects completed a survey after performing all tasks with each selection method. This survey consisted of the NASA TLX questionnaire ~\cite{stanton2013human}. The survey also asked the subjects about their experience with each of the task within the study. The questions were: ``How often did the system select the item you intended?'', ``How often did the system select an item you did not intend to select?'', and ``How quickly was the system able to select your choice.'' All these choices were answered on a scale from zero to five in the increasing order of perceived frequency. They also completed a post-study survey that involved ranking each of the methods for each task.


\subsubsection{Results}


\paragraph{Quantitative Measures} We used a linear mixed effects model\,(\textit{lme4} package in R) to examine whether the interaction between method, task, and new/returning user influenced the measured performance metrics. First, the time-to-completion\,(TTC) was modeled with the following Wilkinson notation: $TTC \sim task*method*returning\_user + (1|user)$, where the task included only the Fitt's Law-style and Sliding Puzzle tasks. The \# of Selections was modeled with the following Wilkinson notation: $\#\,of\,Selections \sim task*method*returning\_user + (1|user) $, where task included only the two Arithmetic Tasks. To probe pairwise tests from significant interactions, we used the \textit{emmeans} package from R. Here, we applied false detection rate correction to reduce the likelihood that a statistical test would produce a significant value by chance. 
No significant interactions or main effects were found for the TTC metric so no further testing was performed on this metric. For selection count, there was a significant three-way interaction\,(\textit{$\beta$} = -127, \textit{p} = 0.004) as well as significant interactions between returning and task\,(\textit{p}s < 0.05) and returning and method\,(\textit{p}s < 0.05). There were main effects within method\,(\textit{p}s < 0.05),\,task (\textit{p}s < 0.05), and returning users\,(\textit{p}s < 0.05). 



\textit{Methods:} GI-G\,($\mu$=192 selections, $p < 0.001$) and GI-P\,($\mu$=181 selections, $p < 0.001$) resulted in significantly more selections than S+I\,($\mu$=46 selections) and Baseline\,($\mu$=69 selections) for returning users in sliding puzzle tasks. 
The pattern of significance was the following: GI-G = GI-P > Baseline = S+I. 
No other pairwise relationships were found.

\textit{Task:} There was a significant difference in the \# of Selections between tasks\,(\textit{p}s < 0.005). This results from the Fitt's Law-style task requiring selection as fast as possible while the puzzle required the user to determine their next move. 

\textit{New/Returning Users:} There was a significant difference in the \# of Selections between returning\,($\mu$=88) and new users\,($\mu$=52) for the Sliding Puzzle task\,(\textit{p}s < 0.005). No other significant relationships were found between user groups.

%

\paragraph{Subjective Measures}
\begin{figure}[h]
    \centering
    \includegraphics[width=0.70\textwidth]{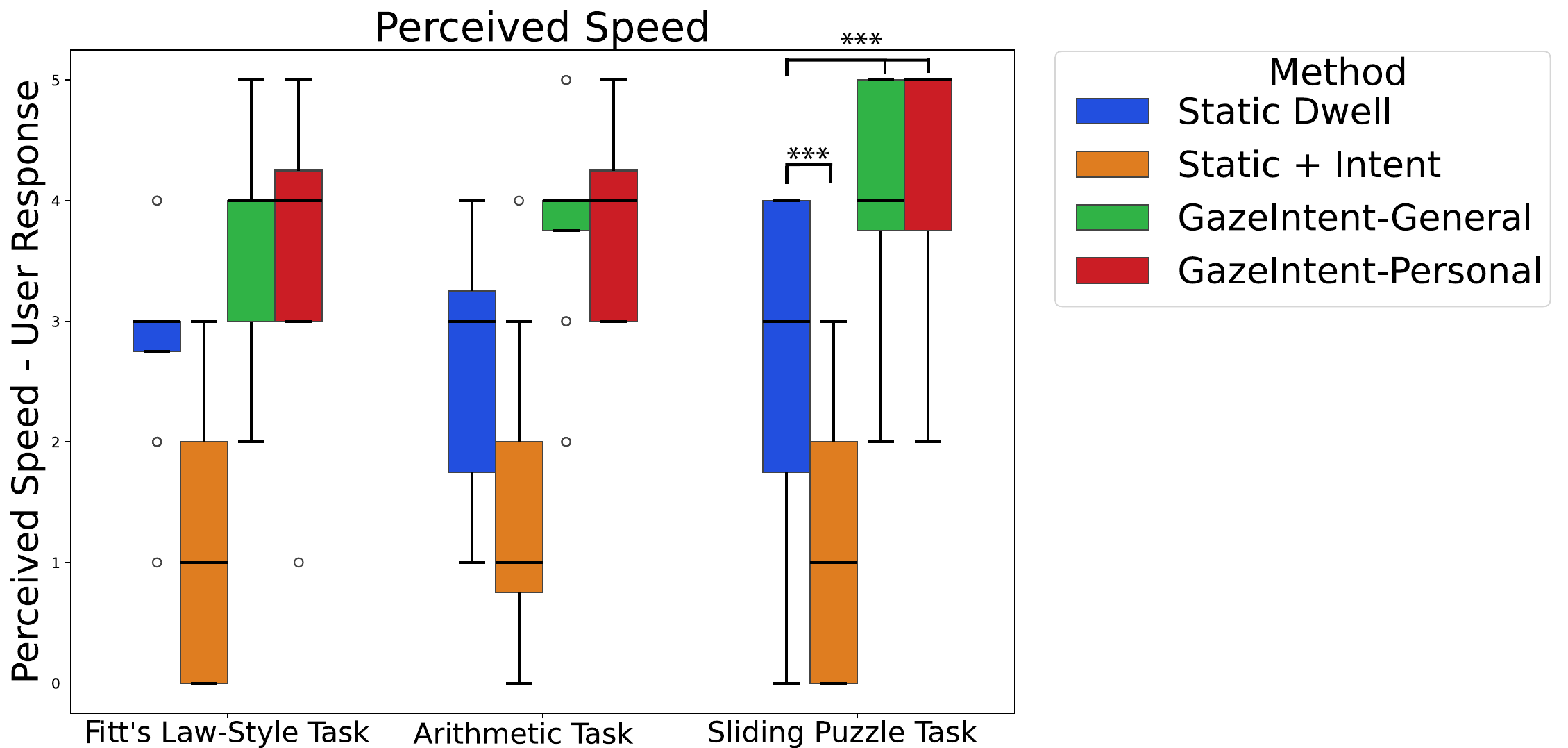}
    \caption{Qualitative Measures: Box plot distributions of subjective speed perceptions across tasks. Users indicated faster system selection with \textit{GazeIntent}-Personal and \textit{GazeIntent}-General. Significantly different groups are marked with *** as \textit{p} < .001.}
    \label{fig:results_qualitative}
\end{figure}

We also applied a linear mixed effects model to examine whether the interaction between task, method, and new/returning user influenced the subjective participant responses from Sec.\,\ref{sec:qual_metrics}\,(Wilkinson notation: $subjective\_responses \sim condition*method*returning\_user + (1|user).$). 
The three-way interactions for subjective false positives and accuracy of selections were not significant\,(all \textit{p}s > 0.05). There was no significant three-way interaction for perceived speed of interaction\,(\textit{p} > 0.05), however, the two-way interaction of method and returning users approached significance\,(\textit{p} = 0.055). 

\begin{figure}[ht]
    \centering
    \includegraphics[width=\textwidth]{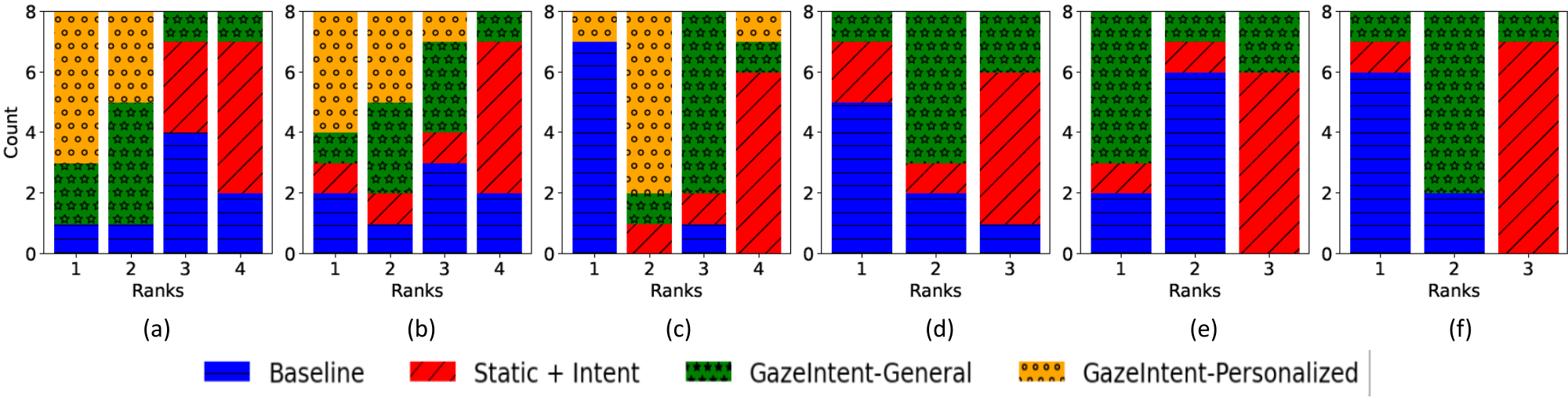}
    \caption{Session-2 User-Rankings, Returning Users (a,b,c) evaluated using four methods and New Users (d,e,f) evaluated using three methods. (a, d): Fitt's Law-like task, (b, e): Arithmetic Task, and (c, f): Sliding Puzzle Task. The rankings indicate that returning users preferred personalized models for Fitt's Law-style and Arithmetic tasks and the Static method was preferred for Sliding Puzzle. New users preferred the generalized GazeIntent method only for the Arithmetic task that it was trained on, and Static baseline otherwise.}
    \label{fig:results_rankings}
\end{figure}

\textit{Speed:} With further probing there were no significant three-way interactions for subjective speed\,(\textit{p} = 0.05512). However, there was a two-way interaction between the method and returning users on the perceived speed of interaction. There was a significant difference between Baseline and GI-G\,(\textit{p} < 0.0001), Baseline and GI-P\,(\textit{p} < 0.0001), Baseline and S + I\,(\textit{p} < 0.0001), S + I and GI-G\,(\textit{p} < 0.0001), and for S + I and GI-P\,(\textit{p} < 0.0001). For the new users, the pairwise interactions between Baseline and GI-G, Baseline and S+I, and S+I and GI-G were significant \,(\textit{p}s < 0.0001). The pattern of significance for returning users was: GI-G = GI-P > Baseline > S+I. For new users, users felt that the \textit{GazeIntent} model was fastest: GI-G > Baseline > S + I. These results show that personalized models were perceived the same as the general model and that \textit{GazeIntent} was perceived as the fastest selection method overall. Visualizations of results for the Subjective and Quantitative measures are provided in the Supplementary Material respectively.
\paragraph{Rankings} Subjects ranked their preferences for each method and each task. Figure~\ref{fig:results_rankings} illustrates the ranking order of methods provided at the end of the experiment segmented by new and returning users.

\textit{Returning Users:} GI-P was the top-ranked method for the Fitt's Law-style task\, (five out of eight participants) and the Arithmetic tasks\,(four out of eight participants). The Static Dwell Baseline was the most preferred method for the sliding puzzle task (seven out of eight people). GI-P was the second most preferred method for the Sliding Puzzle task. Across all of the tasks, S + I was the least preferred method.

\textit{New Users:} Five of eight subjects preferred GI-G for the Arithmetic task, while for the other two tasks the Static Dwell Baseline was preferred. However, GI-G was the second most preferred method in those tasks. Across all tasks, the S+I method was the least preferred method for new users as well. 

\section{Discussion}

Our research goals were to (1) build a comprehensive training dataset for modeling selection intent in VR while varying cognitive load, (2) develop a real-time intent model to drive our \textit{GazeIntent} adaptive dwell time selection method, and evaluate \textit{GazeIntent} against a baseline and state-of-the-art intent-based selection method. We developed the \textit{GazeIntent} adaptive dwell time scaling approach to achieve a balance between false positives and slow selection when requiring a combination of static dwell time and intent model classification.  


\paragraph{Training Dataset} We conducted an initial user study\,(N=16) that provided interaction data using a hand controller while participants performed an arithmetic task. The difficulty of the task and the presence of a memory recall task was used to determine if there was any change in model behavior between conditions. The LSTM intent models were not impacted by changes in cognitive load. We also determined that event-based features did not predict as well as continuous features, which enabled the use of a lightweight intent model using only eleven gaze features that runs in real time on a VR headset. Our trained models outperformed prior results\,(F1 = 0.94).

\paragraph{Model Personalization} The performance of returning users on the same task between GI-G and GI-P indicates the effectiveness of model personalization. Five returning users preferred \textit{GazeIntent} methods and four of those five preferred the model personalized to them. One participant stated that, \textit{``I felt that the first method\,}\,(GI-P)\textit{ provided the most natural feeling selection interaction}.\textit{''} 

\paragraph{Task Generalization}  

A generalizable system will effectively scale dwell time for previously unseen tasks. We measured task generalizability of the \textit{GazeIntent} model trained on Arithmetic data as the number of user selections made in the Circle Selection and the Sliding Puzzle tasks. These two tasks differ from the training data and produce more frequent selections. For the Circle Selection task, the selection count for GI-G was 56\% and 273\% higher than the baseline and S+I methods, respectively. For the Sliding Puzzle task, the GI-G method had 94\% and 177\% higher number of selections compared to the baseline and the S+I method, respectively. These results suggest the deployment of \textit{GazeIntent} scaled dwell times can benefit the speed of interactions even when trained on a different task. 
This indicates that the GazeIntent model was effective within previously unseen tasks. 

\begin{figure}[ht]
    \centering
    \includegraphics[width=0.75\textwidth]{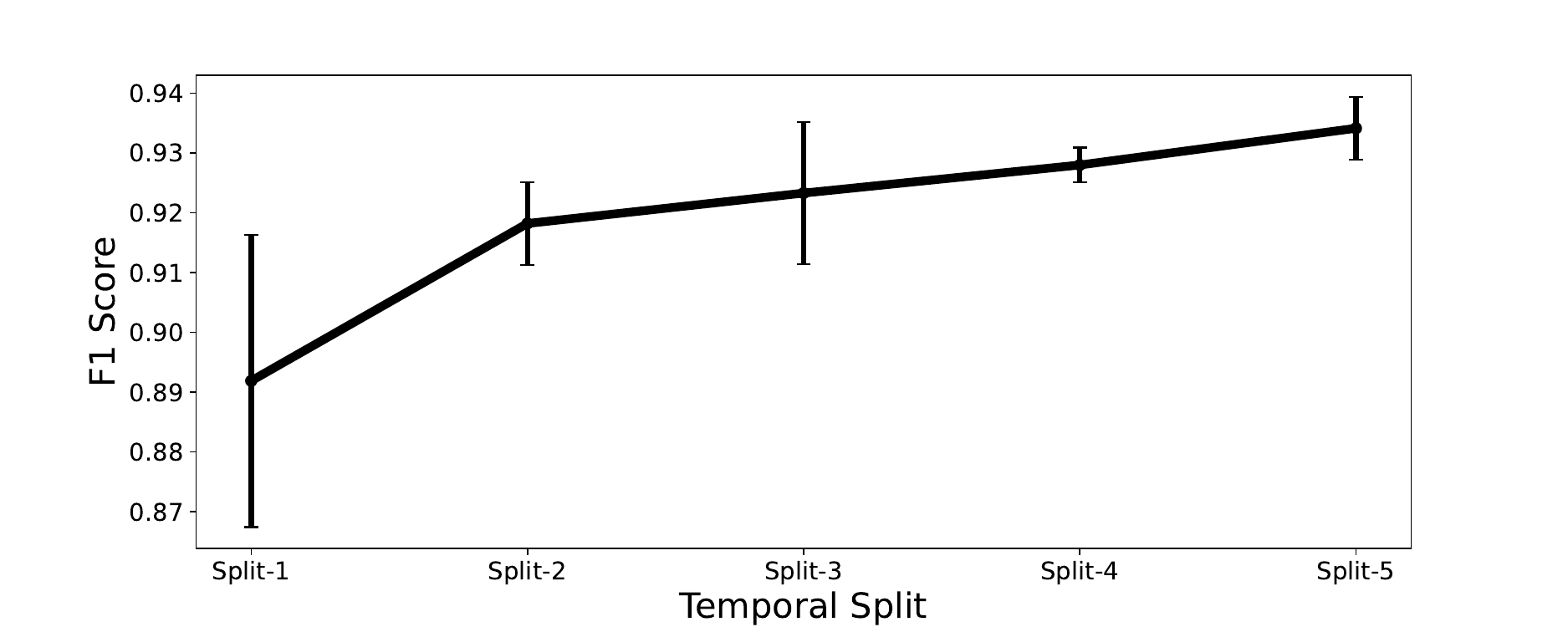}
    \caption{The mean and standard deviation of F1 score for the ten optimal model configurations tested against the entire test dataset. The models are trained on successive temporal splits of training data sessions. Splits 1, 2, 3, 4, and 5 are defined as the models trained sequentially on the first 20\%, 40\%, 60\%, 80\%, and 100\% of the training data. }
    \label{fig:practice_effects}
\end{figure}

\paragraph{Practice Effects} Practice effects refer to modifications in user behavior over prolonged usage of the system. As shown in Fig.\,\ref{fig:practice_effects}, the F1 score of the model increases and has less variability in performance as it is trained successively on user data over time. This shows that our proposed model is learning aspects of the modified behavior of the users over time. 
Qualitatively, 62\% of the returning users preferred \textit{GazeIntent} (GI-G or GI-P) over the other methods on the Division tasks similar to the training data\,(Fig. \,\ref{fig:results_rankings}); suggesting that the deployed model was effective from the start for returning users performing divisibility tests on large numbers again.

\subsection{Limitations}
Our training dataset and evaluation had several limitations due to the scope of our research. Both the data collection and the end user study were performed on a small number of subjects\,(N = 25) between the ages of 20 and 35 who were all right-handed. This affects the generalizability of our findings to larger and more diverse groups of users. The training dataset was collected using controller selection but was deployed in gaze-based selection. While we selected a ground-truth label duration of one second, each individual selection may have had a different lead time between the user deciding they were ready to make the selection and the physical movement of their hand to make the selection. Future work could consider labeling ground truth based on the recorded hand movement instead. We also varied cognitive load in a very specific manner, by using harder math problems and a memory task. Standardized cognitive load assessments or sensors during data collection could account for different levels of cognitive load across users. Although our evaluation results for \textit{GazeIntent} are specific to our LSTM intent model, the dwell time scaling architecture can accommodate any intent prediction model that outputs values between zero and one. 
We believe our primary limitation is how our intent model fused continuous and discrete features. Prior intent modeling work has also struggled to effectively integrate these two streams of gaze-based features~\cite{david2021towards,peacock2022gaze_wm,GazeIntent}. Novel ML models should be explored to best leverage both sets of features, including designing independent models with prediction scores that both contribute to the scaling factor.
 
\subsection{Future Work}
Our approach uses static thresholds ($T_{1}$, $T_{2}$, $T_{3}$, and $T_{4}$) for weighting model predictions based on a pilot study. 
We see the potential to consider adaptation of these thresholds for more effective selection, either to account for temporal practice effects during deployment or to capture changes in gaze behavior between tasks. Adapting these thresholds directly is less intensive than re-training or even fine-tuning the intent model.  The F1 score of the personalized intent models was marginally lower than the generalized model in an offline evaluation. However, most of the returning users preferred the personalized model over the general model in the end-user study. Future work exploring the difference between perceived performance and typical ML metrics requires additional investigation outside of the scope of our paper. It was interesting to observe that the F1 score of the personalized intent models was marginally lower than the generalized model in an offline evaluation. However, most of the returning users indicated that they preferred the personalized model over the general model in the end-user study. Future work exploring the difference between perceived performance and offline ML metrics for interaction requires additional investigation outside of the scope of our paper.


We identify several clear directions for future work to improve the generalizability and stable performance of \textit{GazeIntent}-based interaction. First, the model could be evaluated in free-viewing environments with additional head movements and locomotion. The model could also be evaluated in new interaction environments, including deployment on mixed-reality devices where eye movement behavior may differ between the selection of virtual and physical objects. Additionally, the impact of frequent task switching could affect the performance of gaze-based models. Optimizing threshold constants $T_{1}$, $T_{2}$, $T_{3}$, and $T_{4}$ 
using reinforcement learning or other methodologies would provide a framework for user generalization at deployment. We showed that walk-forward validation successfully adapts to temporal changes in user behavior on a short-term basis. We propose the utility of continual learning ~\cite{de2021continual} 
 to model non-linear practice effects and individual differences between users that could only be studied across long-term or longitudinal data collection. Beyond dwell-based selection, our approach can be extended to gesture-based selection, such as gaze and pinch~\cite{gazegesture2}, to adapt detection thresholds when considering inconsistent gesture recognition and noisy sensor data. 

Finally, GazeIntent was evaluated for tasks with only visual stimuli. Including audio cues has an effect on the prediction of head and eye movements while watching a 360$^{\circ}$ videos~\cite{higuchi2022study}. A study comparing the performance of GazeIntent on various combinations of stimuli, such as audio and tactile cues, could be used to adapt the model across a diverse set of tasks.





\subsection{Privacy \& Ethics}
Considerations of our data collection did not exceed the standards of our IRB  protocol review and secure data management plan. However, privacy and ethics concerns are critical for eye tracking and VR research~\cite{bozkir2023eye}. For example, our intent model could be trained to identify when users are stressed or vulnerable and integrate this inference within a dark or deceptive pattern~\cite{krauss2022exploring,wang2023dark,ramirez2024deceptive}. Similiar ethical discussions are emerging regarding brain-computer interfaces as well~\cite{kablo2023privacy}. 
These implications are true for many predictive models of human behavior and are not unique to our model, and what is considered an ethical application should be addressed through the VR platform~\cite{happa2021privacy}, effective user consent~\cite{selinger2eye}, and any governing frameworks or privacy policies~\cite{spivackRiskFramework}. While implementing and evaluating a privacy-preserving system is out of scope for this paper, we note that prior work has defined safeguards with formal privacy guarantees that defend against re-identification when releasing the training data~\cite{david2023privacy}, when training the model~\cite{abadi2016deep}, and when adding privacy noise to features used as input to the model~\cite{steil2019privacy,bozkir2021differential,david2022your}. 

\section{Conclusion}
Interaction methods for VR benefit from predictive intent models, particularly for gaze-based selection subject to the Midas Touch problem. Our results suggest that intent models can be further used to adapt interaction parameters from existing methods, i.e., dwell-time thresholds, in a manner that accounts for prior data from a user and experience with a task. Our system \textit{GazeIntent} applies a set of thresholds to temporal model predictions in a manner that enables effective interaction across different tasks without the need to re-train the underlying intent model. The thresholds could be adapted by factors beyond task such as individual preferences and practice effects to further improve performance and generalization. Our research advances gaze-based selection using intent models and charts a path toward practical adaptive interfaces for VR and AR applications.      

\begin{acks}
The authors would like to thank Nick Gazzillo for assisting with user studies during the offline training dataset collection, and our anonymous reviewers for their insightful comments and feedback that allowed us to further improve our paper during review.
\end{acks}
\bibliographystyle{ACM-Reference-Format}
\bibliography{citations}
\end{document}